\documentclass[prd,preprint,nofootinbib]{revtex4}

\usepackage{graphicx}
\usepackage{amssymb}
\usepackage{amsmath}

\begin{document}

\renewcommand{\thefootnote}{\#\arabic{footnote}}
\newcommand{\rem}[1]{{\bf [#1]}}
\newcommand{\gsim}{ \mathop{}_ {\textstyle \sim}^{\textstyle >} }
\newcommand{\lsim}{ \mathop{}_ {\textstyle \sim}^{\textstyle <} }
\newcommand{\vev}[1]{ \left\langle {#1}  \right\rangle }
\newcommand{\bear}{\begin{array}}  
\newcommand {\eear}{\end{array}}
\newcommand{\bea}{\begin{eqnarray}}   
\newcommand{\eea}{\end{eqnarray}}
\newcommand{\beq}{\begin{equation}}   
\newcommand{\eeq}{\end{equation}}
\newcommand{\bef}{\begin{figure}}  
\newcommand {\eef}{\end{figure}}
\newcommand{\bec}{\begin{center}} 
\newcommand {\eec}{\end{center}}
\newcommand{\non}{\nonumber}  
\newcommand {\eqn}[1]{\beq {#1}\eeq}
\newcommand{\la}{\left\langle}  
\newcommand{\ra}{\right\rangle}
\newcommand{\ds}{\displaystyle}

\def\SEC#1{Sec.~\ref{#1}}
\def\FIG#1{Fig.~\ref{#1}}
\def\EQ#1{Eq.~(\ref{#1})}
\def\EQS#1{Eqs.~(\ref{#1})}
\def\lrf#1#2{ \left(\frac{#1}{#2}\right)}
\def\lrfp#1#2#3{ \left(\frac{#1}{#2} \right)^{#3}}
\def\GEV#1{10^{#1}{\rm\,GeV}}
\def\MEV#1{10^{#1}{\rm\,MeV}}
\def\KEV#1{10^{#1}{\rm\,keV}}
\def\CO#1{{\cal O}\left({#1}\right)}

\def\lrf#1#2{ \left(\frac{#1}{#2}\right)}
\def\lrfp#1#2#3{ \left(\frac{#1}{#2} \right)^{#3}}

\begin{flushright}
IPMU 09-0005
\end{flushright}

\title{Decaying gravitino dark matter and an upper bound on the gluino mass}

\author{Koichi Hamaguchi${}^{1,2}$,  Fuminobu Takahashi$^{2}$ and T. T. Yanagida${}^{1,2}$}

\affiliation{
${}^{1}$Department of Physics, University of Tokyo, Tokyo 113-0033, Japan,\\
${}^{2}$Institute for the Physics and Mathematics of the Universe,
University of Tokyo, Chiba 277-8568, Japan,
}

\date{\today}

\begin{abstract}
  We show that, if decaying gravitino dark matter is responsible for
  the PAMELA and ATIC/PPB-BETS anomalies in the cosmic-ray electron
  and positron fluxes, both a reheating temperature and a gluino mass
  are constrained from above. In particular, the gluino mass is likely
  within the reach of LHC, if the observed baryon asymmetry is
  explained by thermal leptogenesis scenario.
\end{abstract}


\maketitle


The PAMELA experiment~\cite{Adriani:2008zr} reported an excess of the
positron fraction above $10$ GeV, which extends up to about
$100$\,GeV. The excess could be a signal of the annihilation or decay
of dark mater.  Among many decaying dark matter
models~\cite{Takayama:2000uz,hidden-gauge,composite,misc}, we consider
here the gravitino dark matter with broken
$R$-parity~\cite{Takayama:2000uz} (see also
Refs.~\cite{Ibarra:2007wg}).  In fact, it was shown in
Ref.~\cite{Ishiwata:2008cv} that the gravitino decaying via the
bilinear $R$-parity violation can explain the PAMELA data.

The positron spectrum needed to explain the PAMELA excess is rather
hard.  If the positron fraction continues to rise above $100$\,GeV,
the cosmic-ray electron flux as well may be significantly modified at
high energies.  Interestingly enough, the ATIC balloon experiment
collaboration~\cite{ATIC} has recently released the data, showing a
clear excess in the total flux of electrons plus positrons peaked
around $600 - 700$\,GeV, in consistent with the PPB-BETS
observation~\cite{Torii:2008xu}.  It is highly suggestive of the same
origin for the PAMELA and ATIC/PPB-BETS anomalies, if both are to be
accounted for by dark matter.  As will be shown in Appendix, the
decaying gravitino scenario can actually account for both excesses. We
focus on this scenario in this letter.

The gravitino is assumed to be the lightest supersymmetric particle
(LSP). In the presence of the $R$-parity violation, its longevity is
due to a combination of the Planck-suppressed interactions and a tiny
$R$-parity violating coupling. For the latter we assume the so-called
bilinear $R$-parity violating coupling, which is parametrized by a
dimensionless coupling $\kappa_i$ defined as the ratio of the vacuum
expectation values (VEVs) of the sneutrinos to that of the
standard-model like Higgs boson, where the subindex $i (= 1,2, 3)$
denotes the flavor dependence, $e$, $\mu$ and $\tau$ (see
Ref.~\cite{Ishiwata:2008cu} for more details).  We assume that the
decay of an electron-type dominates over the others throughout this
letter, i.e., $\kappa_1 \gg \kappa_2, \kappa_3$, since one cannot fit well
the sharp cut-off in the ATIC data otherwise.  Then the mass and lifetime of
the gravitino should be in the following range to account for the
PAMELA and ATIC/PPB-BETS excesses:
\bea
\label{mass}
m_{3/2}  &\simeq&  (1.2  -  1.4 ){\rm \,TeV},\\
\label{lifetime}
\tau_{3/2} &\simeq& \CO{10^{26}} {\rm\,sec.}
\eea
Since all the other supersymmetric (SUSY) particles must be heavier
than the gravitino, we expect a typical mass scale for the SUSY
particles, especially the gluino, may be out of the reach of LHC.
This would be quite discouraging for those who expect SUSY discovery
at LHC.  In this letter, however, we show that the gluino mass is
bounded from above and is likely within the reach of LHC, if the
baryon asymmetry is explained by the thermal leptogenesis scenario.

\begin{figure}[t]
\includegraphics[scale=1.0]{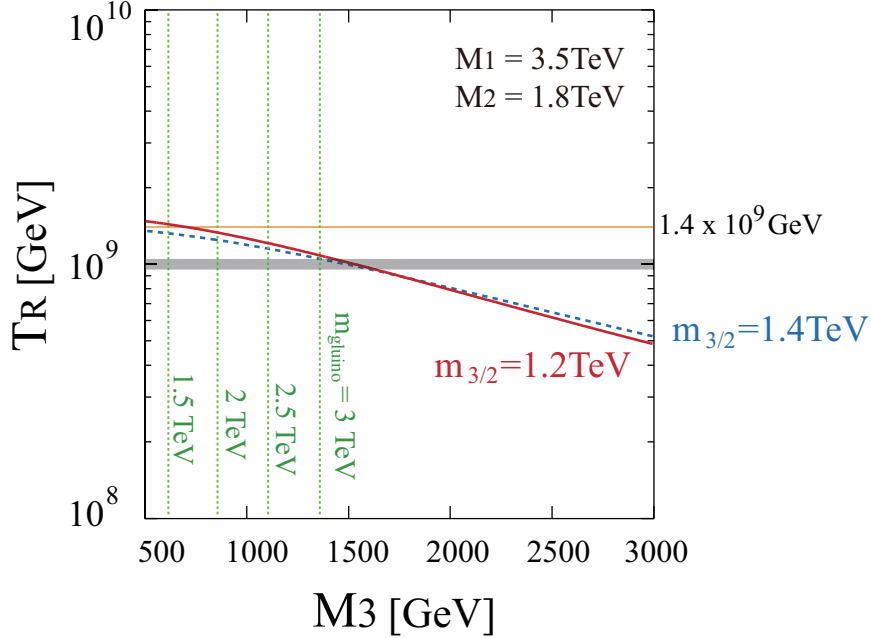}
\caption{The upper bounds on the gluino mass $M_3$ and the reheating
  temperature $T_R$, for the gravitino mass $m_{3/2} = 1.2$\,TeV
  (solid red) and $1.4$\,TeV (dashed blue).  The horizontal thick gray (thin orange) line
  shows the lower bound on $T_R \gtrsim 10^9 (1.4 \times 10^9) $\,GeV for the thermal
  leptogenesis to work.  Here we set the bino and wino masses to be
  $M_1 = 3.5$\,TeV and $M_2 = 1.8$\,TeV.  
  We also show the gluino mass in the low energy, $m_{\rm gluino} = 1.5, 2.0, 2.5, 3.0$\,TeV,
  as the vertical dotted (green) lines from left to right .
  }
\label{fig:gravitino}
\end{figure}

Let us first discuss the gravitino production in the early universe.
The gravitino is produced by thermal scatterings,\footnote{ The
  inflaton decay may also contribute to the gravitino
  abundance~\cite{Kawasaki:2006gs,Takahashi:2007tz}.  We focus on the
  thermal production, since the non-thermal gravitino production
  depends on the inflation models.} and the abundance is given
by~\cite{Moroi:1993mb,Bolz:1998ek,Bolz:2000fu}
\bea
Y_{3/2} &\sim& 4 \times 10^{-12}\, g_3^2(T_R) \,  \ln\lrf{1.3}{g_3(T_R)} \left(1+\frac{M_3^2(T_R)}{3 m_{3/2}^2} \right)
 					 \lrf{T_R}{10^{10}{\rm\,GeV}},
\label{y32}					 
\eea
where $g_3$ and $M_3$ are the $SU(3)_C$ gauge coupling and the gluino
mass, respectively, and both are evaluated at a scale equal to the
reheating temperature, $T_R$, in \EQ{y32}.  For simplicity we have
dropped contributions involving the $U(1)_Y$ and $SU(2)_L$ gauge
interactions, which are subdominant unless the bino and wino masses,
$M_1$ and $M_2$, are much larger than $M_3$.  Thus, the reheating
temperature and the gluino mass are constrained from above for the
gravitino abundance not to exceed the observed dark matter abundance,
$\Omega_{DM}h^2 \simeq 0.1143\pm0.0034$~\cite{Komatsu:2008hk}.  In
Fig.~\ref{fig:gravitino} we have shown the upper bound on the gluino
mass and the reheating temperature, where we have included
contributions from $U(1)_Y$ and $SU(2)_L$ neglected in \EQ{y32}.
We have imposed a requirement that the gravitino abundance should not exceed
the 95\% C.L. upper bound on the dark matter abundance.
We used the code SuSpect2.41~\cite{Djouadi:2002ze} to calculate the gravitino abundance
and the physical spectra for the superparticles, with the following boundary conditions at the
GUT scale $\simeq 2 \times 10^{16}$\,GeV; $\tan \beta = 10$,
${\rm sgn}[\mu] > 0$, the vanishing $A$-terms, the universal scalar mass $m_0 = 2$ TeV 
for the squarks and sleptons, $m_{H_u}^2 = m_{H_d}^2 = 5\times 10^5{\rm GeV}^2$, and the $U(1)_Y$ and
$SU(2)_L$ gaugino masses $M_1 = 3.5$\,TeV and $M_2 = 1.8$\,TeV.
Those parameters are chosen so that the gravitino is LSP~\footnote{
In the case of $m_{3/2} = 1.4$\,TeV, the gravitino is LSP for $M_3 \gtrsim 600$\,GeV
for the adopted parameters. This does not affect the following discussion.
}

The origin of the baryon asymmetry is a big mystery of the modern
cosmology.  The thermal leptogenesis scenario~\cite{Fukugita:1986hr}
is appealing, and the reheating temperature must be higher than about
$2\times10^9$\,GeV~\cite{Buchmuller:2004nz} for the scenario to
work. The precise value of the lower limit depends on flavor
effects~\cite{Blanchet:2006be} and the mass spectrum of the
right-handed neutrinos.  The detailed study showed the lower bound as
$T_R \gtrsim 10^9$ GeV, which is represented by the horizontal gray band in
Fig.~\ref{fig:gravitino}.  We can see from Fig.~\ref{fig:gravitino}
that the gluino mass is bounded from above, $M_3 \lesssim 1.5$\,TeV at
the GUT scale, for $T_R$ to satisfy the lower bound~\footnote{
The upper bound on the gluino mass was also discussed in Refs.~\cite{Fujii:2003nr,Buchmuller:2008vw}
in different  contexts. 
}.  This constraint
can be translated into that the gluino mass should be lighter than
about $3$\,TeV in the low energy, taking account of the
renormalization group evolution.  If we take a slightly tighter bound
on $T_R$, say, $T_R \gtrsim 1.4 \times 10^9$\,GeV, for which the
leptogenesis becomes easier, the gluino mass in the low energy must be
lighter than about $2$\,TeV for $m_{3/2} = 1.2$\,TeV.  This is a surprising result. If the ATIC
anomaly is to be explained by the decay of the gravitino dark matter,
we may worry that the SUSY particles are so heavy that they may not be
produced at LHC.  However, if we believe in the thermal leptogenesis
scenario and impose the lower bound $T_R \gtrsim 1.4(1) \times
10^9$\,GeV, the gluino mass turned out to be lighter than about $2(3)$\,TeV.  
This is a good new for those who anticipate the LHC to
discover SUSY.

Several comments are in order. In the presence of the $R$-parity
violation, it is quite non-trivial whether the SUSY particles can be
detected at LHC, even if they are produced.  If the gluino is the
standard-model LSP, they will escape the detector before it
decays. The collider signature will look like a split SUSY
model~\cite{ArkaniHamed:2004fb}, and it is not easy to collect and
analyze those collider data properly.  On the other hand, if the
lightest SUSY particle is the neutralino, we will observe a large
missing transverse energy. Note that we cannot impose the GUT relation
on the gaugino masses, $M_1 = M_2 = M_3$, since the bino would be
lighter than the gravitino in the low energy.  We have implicitly
assumed that $M_1$ and $M_2$ are not much larger than $M_3$,
throughout this letter. Our argument  will not be significantly modified 
unless $M_1$ and $M_2$ are much larger than $M_3$.

In order to realize the lifetime (\ref{lifetime}) the $\kappa_1$ must
be chosen to be $\kappa_1 \sim 10^{-10}$. Such a tiny $R$-parity
violation can be realized in a scenario that the $R$-parity violation
is tied to the $B-L$ breaking~\cite{Buchmuller:2007ui}.

In this letter we have argued that, if the both PAMELA and
ATIC/PPB-BETS anomalies are accounted for by the decaying gravitino
dark matter, the gluino mass as well as the reheating temperature are
bounded from above. In particular, the gluino is likely well within
the reach of LHC if we assume the thermal leptogenesis
scenario. Unexpected good news from the indirect dark matter
experiments may be indicative of a bright future in the new physics
search at LHC.

\appendix
\section{The decaying gravitino and the PAMELA and ATIC/PPB-BETS excesses}
Here let us show that the decaying gravitino of mass $1.2 - 1.4$\,TeV
can account for both the PAMELA and ATIC/PPB-BETS excesses. For
simplicity we assume the isothermal distribution for dark matter
profile, although our results are not sensitive to the dark matter
profile.  The electron and positron obey the following diffusion
equation,
\begin{equation}
\nabla \cdot\left[K(E,\vec{r})\nabla f_{e}\right]+\frac{\partial}{\partial E}\left[b(E,\vec{r})f_{e}\right]
+Q(E,\vec{r}) \;=\; 0,\,\label{eq:e_prop}
\end{equation}
where $f_{e}$ is the electron number density per unit kinetic energy,
$K(E,\vec{r})$ a diffusion coefficient, $b(E,\vec{r})$ the rate of
energy loss, and $Q(E,\vec{r})$ a source term of the electrons.  The
electron and positron fluxes are related to the number density by
$\Phi = (c/4\pi) f$, where $c$ is the speed of light.  The analytic
solution of \EQ{eq:e_prop} was given in Ref.~\cite{Hisano:2005ec}. In the following
analysis we fix $m_{3/2} = 1.2$\,TeV and $\tau_{3/2} = 1.0 \times
10^{26}$\,sec.  See Ref.~\cite{Delahaye:2007fr} for the values of the
diffusion constant and the energy loss rate, and the details of the
diffusion model parameters.

\begin{figure}[t]
\includegraphics[scale=0.7]{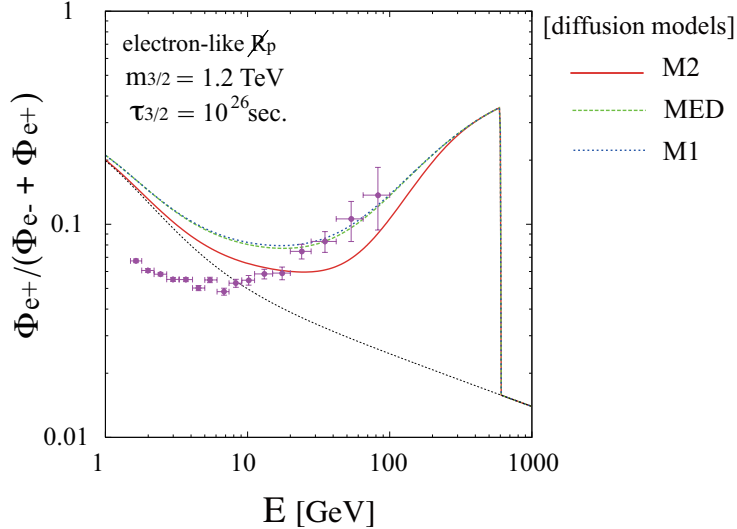}
\caption{The predicted positron fraction for the decaying gravitino
  dark matter, together with the PAMELA data.}
\label{fig:pamela}
\end{figure}

\begin{figure}[t]
\includegraphics[scale=0.7]{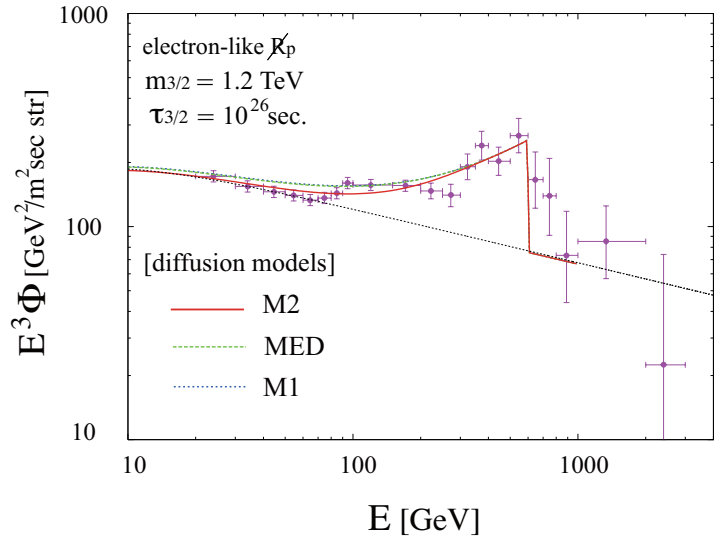}
\caption{The predicted electron plus positron flux for the decaying
  gravitino dark matter, together with the ATIC data.}
\label{fig:atic}
\end{figure}

The bilinear $R$-parity violating operators depend on the lepton
flavor.  We consider the gravitino decay of the electron-type, that
is, $\kappa_1 \gg \kappa_2, \kappa_3$.  In Fig.~\ref{fig:pamela} we
show the positron fraction for the three different diffusion models,
M1, MED and M2, together with the PAMELA data. The MED and M1 models
give a better fit to the data. Similarly we show the predicted
electron plus positron flux together with the ATIC data in
Fig.~\ref{fig:atic}. We have adopted the background for the primary
electrons and the secondary electrons and positrons given in
Ref.~\cite{Moskalenko:1997gh,Baltz:1998xv}, with a normalization
factor $k_{bg} = 0.75$ for the primary electron flux.  As can be seen
from Figs.~\ref{fig:pamela} and \ref{fig:atic}, the gravitino dark
matter can nicely fit both the PAMELA and ATIC data.

For the $R$-parity violating operators of the $\mu-$ and $\tau-$type,
the PAMELA data may be explained, while they give a very poor fit to the ATIC data.

\begin{acknowledgements}
F.T. thanks C-R.Chen for comments.
This work is supported by World Premier International Research Center 
Initiative (WPI Initiative), MEXT, Japan.

\end{acknowledgements}




\begin{thebibliography}{}

\bibitem{Adriani:2008zr}
  O.~Adriani {\it et al.},
  arXiv:0810.4995 [astro-ph].
  
\bibitem{Takayama:2000uz}
  F.~Takayama and M.~Yamaguchi,
  Phys.\ Lett.\  B {\bf 485}, 388 (2000)
  [arXiv:hep-ph/0005214];\\
  W.~Buchmuller, L.~Covi, K.~Hamaguchi, A.~Ibarra and T.~Yanagida,
  JHEP {\bf 0703}, 037 (2007)
  [arXiv:hep-ph/0702184].
  
\bibitem{hidden-gauge}
  C.~R.~Chen, F.~Takahashi and T.~T.~Yanagida,
   Phys.\ Lett.\  B {\bf 671}, 71 (2009)
  [arXiv:0809.0792 [hep-ph]];
  arXiv:0811.0477 [hep-ph];
  C.~R.~Chen, M.~M.~Nojiri, F.~Takahashi and T.~T.~Yanagida,
  arXiv:0811.3357 [astro-ph].


  
\bibitem{composite}  
  K.~Hamaguchi, E.~Nakamura, S.~Shirai and T.~T.~Yanagida,
  arXiv:0811.0737 [hep-ph];
  E.~Nardi, F.~Sannino and A.~Strumia,
  arXiv:0811.4153 [hep-ph];
  K.~Hamaguchi, S.~Shirai and T.~T.~Yanagida,
  arXiv:0812.2374 [hep-ph].
  
  \bibitem{misc}
  C.~R.~Chen and F.~Takahashi,
  arXiv:0810.4110 [hep-ph];
  P.~f.~Yin, Q.~Yuan, J.~Liu, J.~Zhang, X.~j.~Bi and S.~h.~Zhu,
  arXiv:0811.0176 [hep-ph];
  A.~Arvanitaki, S.~Dimopoulos, S.~Dubovsky, P.~W.~Graham, R.~Harnik and S.~Rajendran,
  arXiv:0812.2075 [hep-ph];
  I.~Gogoladze, R.~Khalid, Q.~Shafi and H.~Yuksel,
  arXiv:0901.0923 [hep-ph].

  
  

  
  
\bibitem{Ibarra:2007wg}
  A.~Ibarra and D.~Tran,
  Phys.\ Rev.\ Lett.\  {\bf 100}, 061301 (2008)
  [arXiv:0709.4593 [astro-ph]];
  JCAP {\bf 0807}, 002 (2008)
  [arXiv:0804.4596 [astro-ph]];\\
  K.~Ishiwata, S.~Matsumoto and T.~Moroi,
  arXiv:0805.1133 [hep-ph];\\
  L.~Covi, M.~Grefe, A.~Ibarra and D.~Tran,
  arXiv:0809.5030 [hep-ph].


\bibitem{Ishiwata:2008cv}
  K.~Ishiwata, S.~Matsumoto and T.~Moroi,
  arXiv:0811.0250 [hep-ph].
  

 \bibitem{ATIC}
J.~Chang {\it et al/},
Nature 456 (2008) 362-365.  

\bibitem{Torii:2008xu}
  S.~Torii {\it et al.},
  arXiv:0809.0760 [astro-ph].
 
 
 

\bibitem{Ishiwata:2008cu}
  K.~Ishiwata, S.~Matsumoto and T.~Moroi,
in Ref.~\cite{Ibarra:2007wg}.
 
 
\bibitem{Kawasaki:2006gs}
  M.~Kawasaki, F.~Takahashi and T.~T.~Yanagida,
  Phys.\ Lett.\  B {\bf 638}, 8 (2006)
  [arXiv:hep-ph/0603265];
  Phys.\ Rev.\  D {\bf 74}, 043519 (2006)
  [arXiv:hep-ph/0605297].
  
\bibitem{Takahashi:2007tz}
  F.~Takahashi,
  Phys.\ Lett.\  B {\bf 660}, 100 (2008)
  [arXiv:0705.0579 [hep-ph]].
  
  
\bibitem{Moroi:1993mb}
  T.~Moroi, H.~Murayama and M.~Yamaguchi,
  Phys.\ Lett.\  B {\bf 303}, 289 (1993).

\bibitem{Bolz:1998ek}
  M.~Bolz, W.~Buchmuller and M.~Plumacher,
  Phys.\ Lett.\  B {\bf 443}, 209 (1998)
  [arXiv:hep-ph/9809381].


\bibitem{Bolz:2000fu}
  M.~Bolz, A.~Brandenburg and W.~Buchmuller,
  Nucl.\ Phys.\  B {\bf 606}, 518 (2001)
  [Erratum-ibid.\  B {\bf 790}, 336 (2008)]
  [arXiv:hep-ph/0012052];
  J.~Pradler and F.~D.~Steffen,
  Phys.\ Rev.\  D {\bf 75}, 023509 (2007)
  [arXiv:hep-ph/0608344].


\bibitem{Komatsu:2008hk}
  E.~Komatsu {\it et al.}  [WMAP Collaboration],
  arXiv:0803.0547 [astro-ph].
  
\bibitem{Djouadi:2002ze}
  A.~Djouadi, J.~L.~Kneur and G.~Moultaka,
  Comput.\ Phys.\ Commun.\  {\bf 176}, 426 (2007)
  [arXiv:hep-ph/0211331].
 
  
\bibitem{Fukugita:1986hr}
  M.~Fukugita and T.~Yanagida,
  Phys.\ Lett.\  B {\bf 174}, 45 (1986).
  
\bibitem{Buchmuller:2004nz}
  W.~Buchmuller, P.~Di Bari and M.~Plumacher,
  Annals Phys.\  {\bf 315}, 305 (2005)
  [arXiv:hep-ph/0401240].

\bibitem{Blanchet:2006be}
  S.~Blanchet and P.~Di Bari,
  JCAP {\bf 0703}, 018 (2007)
  [arXiv:hep-ph/0607330];
        S.~Antusch and A.~M.~Teixeira,
        JCAP {\bf 0702}, 024 (2007)
        [arXiv:hep-ph/0611232].

 
\bibitem{Fujii:2003nr}
  M.~Fujii, M.~Ibe and T.~Yanagida,
  Phys.\ Lett.\  B {\bf 579}, 6 (2004)
  [arXiv:hep-ph/0310142].
 
 
\bibitem{Buchmuller:2008vw}
  W.~Buchmuller, M.~Endo and T.~Shindou,
  JHEP {\bf 0811}, 079 (2008)
  [arXiv:0809.4667 [hep-ph]].
 
 
\bibitem{ArkaniHamed:2004fb}
  N.~Arkani-Hamed and S.~Dimopoulos,
  JHEP {\bf 0506}, 073 (2005)
  [arXiv:hep-th/0405159];
  G.~F.~Giudice and A.~Romanino,
  Nucl.\ Phys.\  B {\bf 699}, 65 (2004)
  [Erratum-ibid.\  B {\bf 706}, 65 (2005)]
  [arXiv:hep-ph/0406088].

 
 
\bibitem{Buchmuller:2007ui}
  W.~Buchmuller, L.~Covi, K.~Hamaguchi, A.~Ibarra and T.~Yanagida,
in Ref.~\cite{Takayama:2000uz}.
 
\bibitem{Hisano:2005ec}
  J.~Hisano, S.~Matsumoto, O.~Saito and M.~Senami,
  Phys.\ Rev.\  D {\bf 73}, 055004 (2006)
  [arXiv:hep-ph/0511118].

\bibitem{Delahaye:2007fr}
  T.~Delahaye, R.~Lineros, F.~Donato, N.~Fornengo and P.~Salati,
  Phys.\ Rev.\  D {\bf 77}, 063527 (2008)
  [arXiv:0712.2312 [astro-ph]].
  
\bibitem{Moskalenko:1997gh}
  I.~V.~Moskalenko and A.~W.~Strong,
  Astrophys.\ J.\  {\bf 493}, 694 (1998)
  [arXiv:astro-ph/9710124].
 
\bibitem{Baltz:1998xv}
  E.~A.~Baltz and J.~Edsjo,
  Phys.\ Rev.\  D {\bf 59}, 023511 (1999)
  [arXiv:astro-ph/9808243].

  
\end{thebibliography}
\end{document}